\documentclass[prl,twocolumn,superscriptaddress,showpacs,floatfix,longbibliography]{revtex4-1}
\usepackage{mathrsfs,braket}
\usepackage{amssymb, amsbsy, amsmath, latexsym, dsfont, array, layout, graphicx,mathrsfs,color, bm}
\usepackage[unicode=true,breaklinks=true,colorlinks=true]{hyperref}
\hypersetup{citecolor=blue,urlcolor=blue}

\begin{document}
\title{Scale-free localization in pseudo-Hermitian quantum walk}

\author{Haiting Zhang} \thanks{These authors contributed equally to this work.}
\affiliation{Beijing Computational Science Research Center, Beijing 100193, China}

\author{Lei Xiao} \thanks{These authors contributed equally to this work.}
\affiliation{School of Physics, Southeast University, Nanjing 211189, China}
\affiliation{Key Laboratory of Quantum Materials and Devices of Ministry of Education, Southeast University, Nanjing 211189, China}

\author{Cui-Xian Guo}
\affiliation{Beijing Key Laboratory
	of Optical Detection Technology for Oil and Gas, China University of Petroleum-Beijing, Beijing 102249, China}
\affiliation{Basic Research Center for Energy Interdisciplinary, College of Science, China University of Petroleum-Beijing, Beijing 102249, China}

\author{Quan Lin}
\affiliation{School of Physics, Southeast University, Nanjing 211189, China}
\affiliation{Key Laboratory of Quantum Materials and Devices of Ministry of Education, Southeast University, Nanjing 211189, China}

\author{Peng Xue}\email{gnep.eux@gmail.com}
\affiliation{School of Physics, Southeast University, Nanjing 211189, China}
\affiliation{Key Laboratory of Quantum Materials and Devices of Ministry of Education, Southeast University, Nanjing 211189, China}

%%%%%%%%%%%%%%%%%%%%%%%%%%%%%%%%%%%%%%%%%%%%%%%%%%%%%%%%%%%%%%%%%%%%%%%%%%%%%%%%%%%%%%%%%%%%%%%%%%%%%%%%%%%%%%%%%%%%%%%%
\begin{abstract}
Scale-free localization is a distinctive phenomenon in non-Hermitian systems, where specific eigenstates, known as scale-free localized states (LSs), possess localization lengths that scale linearly with the system size.
Here, we show that  a single non-Hermitian defect in  one-dimensional quantum walks (QWs) induces scale-free localization.
Driven by non-Hermiticity, the QW system undergoes a pseudo-Hermitian (PH) phase transition and the PH symmetry breaking is accompanied by the emergence of scale-free localization.
As non-Hermiticity is further increased, the system undergoes a second phase transition, marked by the emergence of two pairs of bound states.
Crucially, we identify the critical points of both transitions using localization properties.
Furthermore, we extend our investigation to two-dimensional QWs, where a single non-Hermitian defect also induces scale-free localization, thereby demonstrating the universality of this phenomenon.
Our work  provides a valuable perspective for further exploring defect physics and a type of novel LS in non-Hermitian quantum systems.
\end{abstract}
\maketitle
		
\textit{Introduction.---}Over the past decade, non-Hermitian systems~\cite{C93,BB98} have attracted significant attention,
revealing a wealth of phenomena far richer than those observed in Hermitian systems.
These exotic non-Hermitian phenomena have been revealed, such as parity-time symmetry~\cite{KSUS19,ZL19}, pseudo-Hermitian (PH)-symmetry~\cite{WKPLN17}, non-Hermitian skin effect~\cite{GGXLS22,R21,SWY19,LWWLMJ24,SP21,SZH21,LMG20}, exceptional points~\cite{Xue26,MZF18,DMXZC16,PMR22}, non-Bloch bulk-boundary correspondence~\cite{YW18,YSW18}, anomalous
localization transitions~\cite{HF16,FZ99,FZ999,HN96}, and enlarged symmetry classifications of topological matter~\cite{SO21,CWZLN25,S19}.
Notably, a range of non-Hermitian phenomena have been observed on various experimental platforms, including cold atoms~\cite{GCXXD20}, photonics~\cite{XWD+25,XDWWY21,GSDNV09,ZRPLN15}, acoustic systems~\cite{ZFLSL18}, electrical circuits~\cite{HHSSB20}, and nitrogen-vacancy centers in diamonds~\cite{WLGSY19}.
However, numerous unique features of non-Hermitian systems are caused by the sensitivity to boundary conditions~\cite{KEBB18,EA22}.
For instance, the non-Hermitian skin effect stems from the global non-Hermiticity of the system, whereby eigenstates become exponentially localized~\cite{HTHK94,WCY21,L21} at the boundaries under open boundary conditions~\cite{YM19,LC20}.

Previously, most existing studies have primarily focused on non-Hermitian systems characterized by global gain/loss mechanisms, where non-Hermitian terms act uniformly across the entire system.
However, band theory~\cite{BKS20,GAKTH18} indicates that  a single non-Hermitian defect~\cite{MAB23} introduced into a Hermitian system  has a remarkable impact on the properties of the system, such as transport~\cite{YW26}, localization~\cite{WSW23,LWSW23,YYB24}.
In particular, a novel non-Hermitian phenomenon known as scale-free localization has been identified~\cite{LLG21,GWHC23,XLMDP24,WLLJ25}, characterized by scale-free LSs whose spatial profile remains invariant with respect to the system size.
Despite significant theoretical advances,  how scale-free localization relates to multiple phase transitions and to higher dimensions remains largely unexplored.
Therefore, in this context, we propose to study scale-free localization in one-dimensional (1D) and two-dimensional (2D) quantum walks (QWs)~\cite{ADZ93,AO13,OANK15,SK10,OK11,RA15,CK16} with a single non-Hermitian defect.
The advantage of photonic QWs lies not only in easily introducing gain and loss, but also in their natural capability for dimensional extension.

In this work, we construct a 1D QW with PH-symmetry by introducing a single non-Hermitian defect.
By analyzing  the energy spectrum, we demonstrate that the system undergoes  a PH phase transition and a second phase transition associated with the emergence of bound states.
Upon PH-symmetry breaking, some extended states transition into scale-free localized states (LSs).
Furthermore, we show that for a scale-free LS, the IPR scales linearly with the inverse of the system size, whereas for a bound state, the IPR remains constant independent of the system size.
More importantly, using the averaged IPR and the normalized participation ratio (NPR), we identify the critical points of both phase transitions.
Finally, we extend our study by constructing a 2D QW that also exhibits scale-free localization, thereby confirming the universality of this phenomenon.
Our work provides a valuable perspective for the further exploration of non-Hermitian defect physics.

\begin{figure*}[htp!]	
\centering 	
\includegraphics[width=1\linewidth]{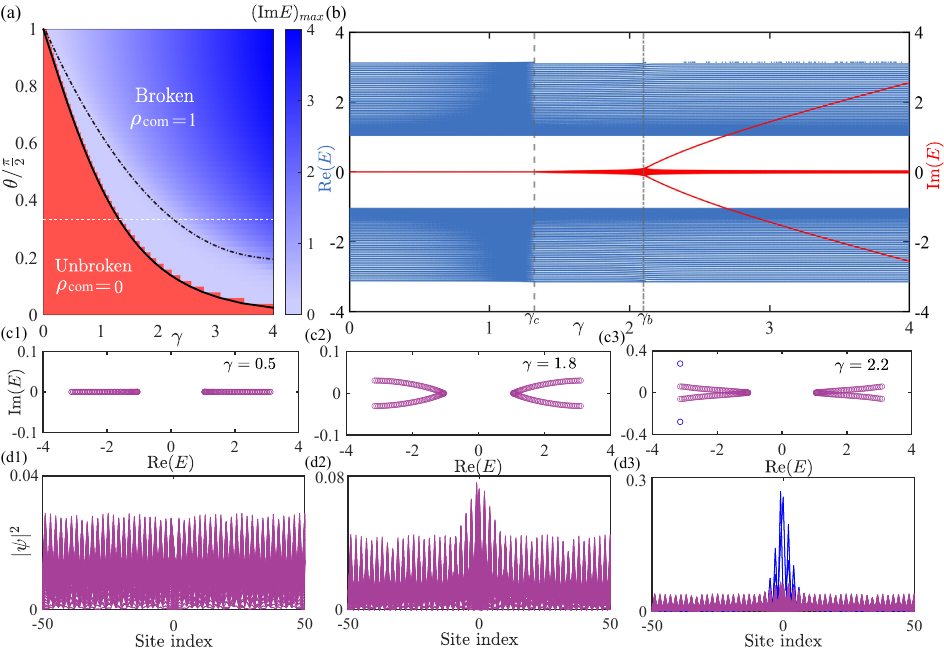}
\caption{\label{fig:1}
  (a) Phase diagram of the QW with $L=101$. The maximal imaginary eigenvalue $(\text{Im}E)_\text{max}$ as a function of the coin parameters $\theta$ and the non-Hermitian strength $\gamma$. The red region is the PH-symmetry unbroken phase  with $\rho_{\text{com}}=0$.
The blue region is the PH-symmetry broken phase with $\rho_{\text{com}}=1$.
The black solid line marks the phase boundary $\gamma_c$ of the PH-symmetry breaking.
The black  dash-dotted line marks another
phase boundary $\gamma_b$ characterized by the emergence of bound states.
(b) The real and imaginary parts of the eigenvalues are plotted in blue and red, respectively, as functions of $\gamma$ for $\theta = \pi/6$. The gray dashed and dash-dotted  lines indicate the critical points $\gamma_c=1.32$ and $\gamma_b=2.09$ of the two phase transitions.
(c1)-(c3) Spectra on the complex-energy plane and (d1)-(d3) Spatial distributions of all eigenstates with $\theta=\pi/6$ for $\gamma=0.5,1.8,2.2$, respectively. In the last panel, the blue lines show the spatial distributions of two pairs of bound states.}
\end{figure*}

\textit{Model.---}We consider a split-step QW \cite{XZBWZ17,WQZBS19,XDWZW20,ZWLX25} on a 1D lattice under periodic boundary conditions, with a single non-Hermitian defect introduced via on-site gain and loss terms, defined by
\begin{equation}
U=D^{-1}SR(\theta)DSR(\theta),
\label{eq:U}		
\end{equation}
where the coin operator is
\begin{equation}
R(\theta)=\sum_x|x\rangle\langle x|\otimes
\begin{pmatrix}
\cos\theta & -\sin\theta \\
\sin\theta & \cos\theta
\end{pmatrix}.
\end{equation}
The conditional position shift operator is given by
\begin{equation}
S=\sum_{x}|x+1\rangle\langle x|\otimes|0\rangle\langle 0| + |x-1\rangle\langle x|\otimes|1\rangle\langle 1|,
\end{equation}
where $x$ denotes the position of the walker and
$\{|1\rangle,|0\rangle\}$ are two orthogonal coin states.
 The single non-Hermitian defect is introduced via $D$ and $D^{-1}$, defined as
\begin{equation}
D = \sum_{x \neq 0} |x\rangle\langle x| \otimes \mathbb{I} + |0\rangle\langle 0| \otimes \begin{pmatrix} e^{\gamma} & 0 \\ 0 & e^{-\gamma} \end{pmatrix},
\end{equation}
where $\mathbb{I}$ is the identity operator and $\gamma$ is the non-Hermitian strength.
Consequently, an effective Hamiltonian $H$ can be derived from  $H=i\ln U$ in Eq.~\eqref{eq:U}.
The eigenvalues of $H$ are given by $E = i\ln\lambda$, where $\lambda$ is the eigenvalue of $U$.

\textit {Pseudo-Hermitian symmetry.---}We find that $U$ satisfies the PH condition~\cite{BB98,BBP98}, i.e., there exists an invertible matrix $\eta$ such that $\eta U^{-1} \eta^{-1}=U^\dagger$ ($\eta H \eta^{-1}=H^\dagger$)~\cite{supp}. This condition is fulfilled by defining $\eta$ as
\begin{equation}
\eta=\sum_n|\tilde\psi_n\rangle\langle\tilde\psi_n|,\label{eq:eta}
\end{equation}
where $|\tilde\psi_\textit{n}\rangle$ is the left eigenstate of $U$.
When the system is in the PH-symmetry unbroken phase, all eigenvalues are real. Conversely, in the PH-symmetry broken phase, paired complex-conjugate eigenvalues emerge.

To characterize these spectral transitions, we introduce the following quantities
\begin{align}
    \rho_{\text{com}}&=P_{\text{com}}/2L,\\
    (\textrm{Im}E)_\text{max}&=\underset{i=1,\dots,2L}{\text{max}}\text{Im}E_i.
\end{align}
Here  $P_{\text{com}}$ represents the number of eigenvalues of $H$ whose imaginary parts are nonzero, and $E_i$ is the $i$th eigenvalue of $H$.
Thus, $\rho_{\text{com}}$ and $(\text{Im}E)_\text{max}$  represent  the ratio of the number of complex eigenvalues and the maximal imaginary part of $E$ over the spectrum at a given set of system parameters.
 Figure~\ref{fig:1}(a) shows the phase diagram in the 1D QW with the maximal imaginary eigenvalue $(\text{Im}E)_\text{max}$ by varying the coin parameters $\theta$ and the non-Hermitian strength $\gamma$, for a sufficiently large system size. 
 In the red region,  we have $\rho_{\text{com}}=0$ and $(\text{Im}E)_\text{max}=0$, indicating the system is PH-symmetry unbroken. In the blue region, the system is PH-symmetry broken with $\rho_{\text{com}}=1$ and $(\text{Im}E)_{max}>0$.
A detailed analysis of $\rho_{\text{com}}$ can be found  in~\cite{supp}.

Using a small number of lattice sites, the PH phase boundary can be analytically expressed as
\begin{equation}
\gamma_c =
\begin{cases}
\displaystyle
\ln\!\left( {\tan\frac{\theta}{2}} \right), & \theta \neq \pi/2 + k\pi, \\
0, & \theta= \pi/2 + k\pi,
\end{cases}
\label{eq:gamma}
\end{equation}
where $k \in \mathbb{Z}$.
 Notably, as marked by the black solid line in Fig.~\ref{fig:1}(a), the analytical expression agrees well with the numerical results.
The system's symmetry is periodic in $\theta$ with period $\pi$. 
Moreover, the critical point of the PH phase transition  corresponds to multiple second-order exceptional points, where 
$\eta$  develops  zero eigenvalues and becomes singular.
Increasing the non-Hermitian strength to $\gamma_b$
  induces another phase transition, namely the emergence of bound states.
 Across the bound state emergence region, $(\text{Im}E)_{max}$ changes sharply. We fit the phase boundary (black dash-dotted line in Fig.~\ref{fig:1}(a)) for large system sizes as
\begin{equation}
\gamma_b\approx 1.92\ln\left(\frac{\pi}{2\theta}\right), \quad \theta\in(0,\pi/2].
\end{equation}
A detailed explanation of $\eta$ and exceptional points can be found  in~\cite{supp}.

As an example, for $\theta=\pi/6$,
 Fig.~\ref{fig:1}(b) shows the real and imaginary parts of the eigenvalues as a function of $\gamma$.
The PH phase transition occurs at $\gamma_c\approx1.32$, which agrees with Eq.~\eqref{eq:gamma},
and another phase transition involving the emergence of bound states occurs at $\gamma_b\approx2.09$.
For $\gamma<\gamma_c$ (PH-symmetry unbroken), all eigenvalues are real [Fig.~\ref{fig:1}(c1)] and all eigenstates are extended [Fig.~\ref{fig:1}(d1)].
When the symmetry is broken with increasing non-Hermitian strength, two distinct regimes emerge, as shown in Figs.~\ref{fig:1}(c2) and (c3).
For $\gamma_c<\gamma<\gamma_b$,  some eigenstates become LSs in Fig.~\ref{fig:1}(d2).
For $\gamma > \gamma_b$, two pairs of bound states appear, and their eigenvalues exhibit significantly larger imaginary parts in Fig.~\ref{fig:1}(d3). 

\textit {Scale-free Localization.---}
So far, we have focused on investigating the spectral properties induced by a single non-Hermitian defect in the 1D QW. To clarify the types of these LSs, we next  analyze their spatial profile and the scaling of the IPR~\cite{LLXWY22,QGZ24} of systems with different sizes.
We first introduce the normalized spatial profile
\begin{equation}
|\Psi_n(x)|^2=\frac{|\psi_n(x)|^2}{\max\left[{|\psi_n(x)|^2}\right]},
\end{equation}
where $|\psi_n(x)|^2$ denotes the original spatial probability density. Additionally, the IPR for a given eigenstate is defined as
\begin{equation}
\textrm{IPR}_n=\sum_x\frac{|\psi_n(x)|^4}{\left(\sum_x|\psi_n(x)|^2\right)^2},	
\end{equation}
where $\psi_n(x)$ represents the $n$th eigenstate of $U$ with $n=1,\dots,L$. A larger IPR value indicates that the state is more highly localized.

\begin{figure}[t!]	
	\centering 	
	\includegraphics[width=1\linewidth]{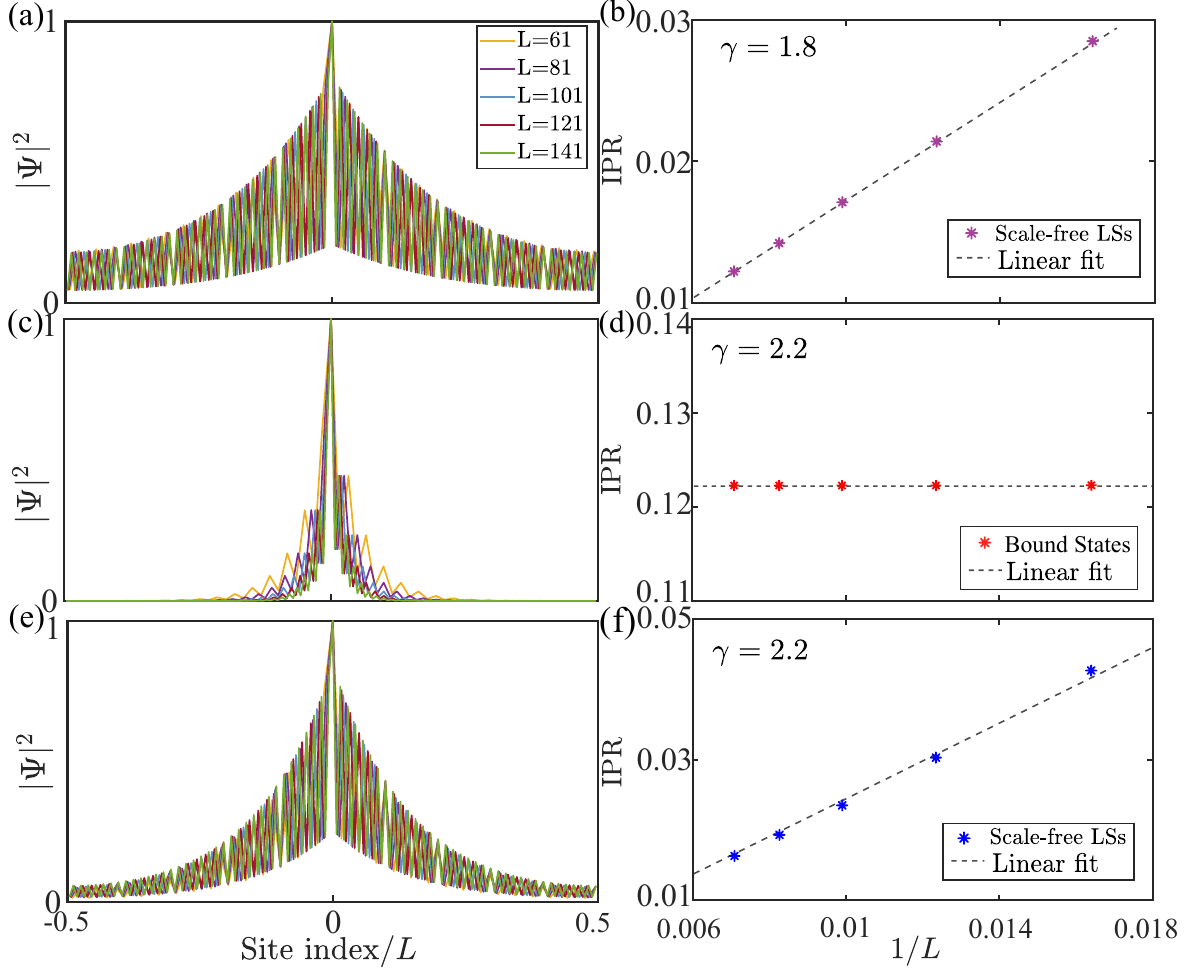}
	\caption{\label{fig:2}
		Left panels:  Rescaled normalized spatial profiles of eigenstates for different system sizes (a) $\gamma=1.8$ (the largest imaginary part), (c) $\gamma=2.2$ (the largest imaginary part), (e)
$\gamma=2.2$ (the third-largest imaginary part).
The horizontal axis is normalized by $L$.
Right panels:  IPR as a function of  $1/L$ corresponding to the left panels (b) scale-free LSs, (d) bound states, (f)
 scale-free LSs.
The black dotted lines denote the linear fit.}
\end{figure}

We present the eigenstates of $H$ for $\gamma = 1.8$ and $2.2$ in the PH-symmetry broken phase,  revealing that scale-free LSs exist at both strengths, whereas bound states appear only at $\gamma = 2.2$.
Figure~\ref{fig:2}(a) shows the normalized spatial profiles of the eigenstates with the largest imaginary eigenvalue for different system sizes at $\gamma = 1.8$.  Evidently, regardless of system size, the spatial profiles of all eigenstates overlap, indicating that these states are scale-free LSs.
As shown in Fig.~\ref{fig:2}(b), the scale-free nature of these states is further quantified by the linear scaling of their IPR with $1/L$.

 In contrast, for $\gamma = 2.2$, the normalized spatial profiles and the scaling of the IPR for the eigenstates with the  imaginary parts of the eigenvalues change significantly across different system sizes, exhibiting characteristics typical of bound states.
As shown in Figs.~\ref{fig:2}(c) and (d), the spatial profiles of these states vary with system size while their IPR remains constant as a function of $1/L$, confirming their bound state nature.
To further indicate scale-free LSs at $\gamma = 2.2$, we examine the LSs with the third-largest imaginary part of the eigenvalues, excluding the bound states themselves for various system sizes.
As shown in Figs.~\ref{fig:2}(e) and (f), these LSs retain similar spatial profiles across different system sizes, and their IPR scales linearly with $1/L$,  a behavior that is consistent with scale-free localization.
The localization lengths~\cite{supp} are obtained by fitting the numerical eigenstates to the spatial profiles, which further confirms the scale-free LSs and bound states.

The above phenomenon clearly indicates that both scale-free LSs and bound states originate from a single non-Hermitian defect in the 1D QW.
As shown in Figs.~\ref{fig:2}(b) and (d), another interesting feature is the significantly higher IPR of the bound states compared to that of the scale-free LSs. Therefore, in addition to identifying the critical points of these two phase transitions from the energy spectrum, we can further elucidate them through their localization characteristics.

\textit {Critical points.---}To quantitatively identify the critical points of these two phase transitions, we  introduce the averaged IPR \cite{CLCW13,WZZY20} and the averaged NPR \cite{LLXWY22,LXL17}, defined respectively as
\begin{equation}
\begin{split}
{\langle \textrm{IPR}\rangle}&=\frac{1}{L}\sum_{n=1}^L\textrm{IPR}_n,	\\
{\langle \textrm{NPR}\rangle}&=\frac{1}{L}\sum_{n=1}^L\textrm{NPR}_n,	
\end{split}
\end{equation}
where $\textrm{NPR}_n={\textrm {IPR}_n}^{-1}/L$.
In the PH-symmetry unbroken phase, all eigenstates are extended. Consequently,  $\langle \mathrm{IPR} \rangle$ assumes a small value, while  $\langle \mathrm{NPR} \rangle$ remains comparatively large.
When the system enters the PH-symmetry broken phase, some eigenstates become scale-free LSs, resulting in an increase in $\langle \mathrm{IPR} \rangle$ and a corresponding decrease in $\langle \mathrm{NPR} \rangle$.
A further increase in the non-Hermitian strength leads to the emergence of bound states.
These highly localized states  further enhance $\langle \mathrm{IPR} \rangle$ and suppress $\langle \mathrm{NPR} \rangle$, marking another phase transition.

\begin{figure}[htp!]	
\centering 	
\includegraphics[width=1\linewidth]{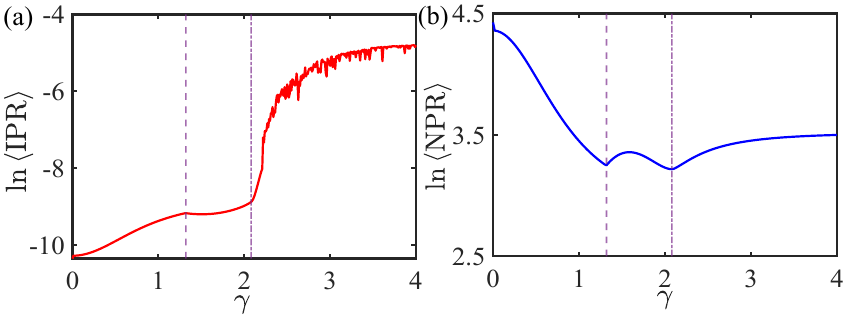}
\caption{\label{fig:3}
(a) Rescaled averaged IPR as a function of $\gamma$ with $L=201$.
(b) Rescaled averaged NPR  as a function of $\gamma$ with $L=201$.
In each panel, the purple dashed and dash-dotted  lines
mark the critical points $\gamma_c=1.32$ and $\gamma_b=2.09$, corresponding to the PH-symmetry breaking transition and the bound-state emergence transition, respectively.
}
\end{figure}

Therefore, we adopt $\ln\langle \mathrm{IPR} \rangle$ and $\ln\langle \mathrm{NPR} \rangle$ as observables to detect the critical points of these two phase transitions.
As shown in Figs.~\ref{fig:3}(a) and (b), both quantities exhibit clear singularities at $\gamma_c \approx 1.32$ and $\gamma_b \approx 2.09$, in agreement with the critical points of the two phase transitions determined from the spectral analysis in Fig.~\ref{fig:1}(b).

\begin{figure*}[htp!]	
\centering 	
\includegraphics[width=1\linewidth]{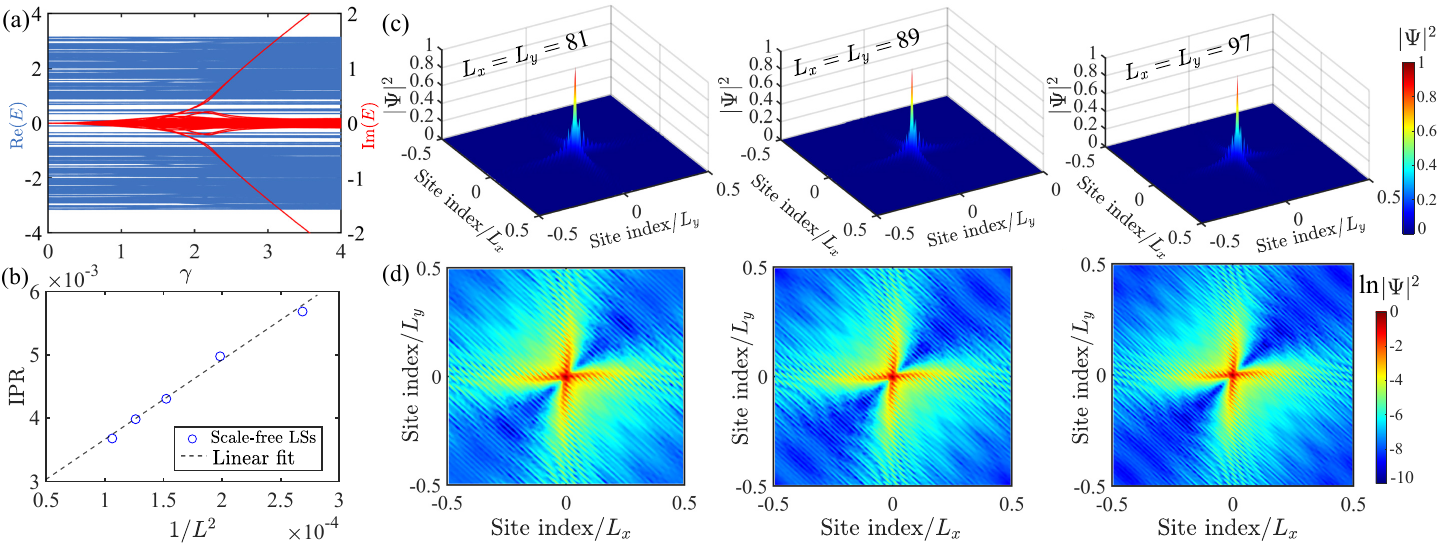}
\caption{\label{fig:4}
(a) The real and imaginary parts of the 2D QW eigenvalues are plotted in blue and red, respectively, as functions of $\gamma$ for $\theta = \pi/6$.
(b) IPR  as a function of $1/L^2$ for scale-free LSs with the fifth-largest imaginary part.
(c) Rescaled normalized spatial profiles
of eigenstates for different system sizes. The horizontal axes are normalized by $L$.
(d) Top views of the profiles shown in (c), plotted on a  logarithmic scale.
}
\end{figure*}

\textit {Scale-free localization in Two-dimensional quantum walks.---}
To illustrate the universality of scale-free localization, we extend our study to a 2D QW with a single non-Hermitian defect.
The 2D split-step QW is defined by
\begin{equation}
U_2=D_2^{-1}S_2R_2(\theta)D_2S_2R_2(\theta),
\label{eq:U2}		
\end{equation}
where the coin operator is
\begin{equation}
R_2(\theta)=\sum_{x,y}|x,y\rangle\langle x,y|\otimes
\begin{pmatrix}
\cos\theta & -\sin\theta \\
\sin\theta & \cos\theta
\end{pmatrix}
\otimes
\begin{pmatrix}
\cos\theta & -\sin\theta \\
\sin\theta & \cos\theta
\end{pmatrix}.
\end{equation}
The conditional position shift
operator is given by
\begin{equation}
\begin{aligned}
S_2&=\sum_{x,y}|x+1,y\rangle\langle x,y|\otimes|00\rangle\langle 00|+ |x-1,y\rangle\langle x,y|\otimes|01\rangle\langle 01| \\
&+ |x,y-1\rangle\langle x,y|\otimes|10\rangle\langle 10|+|x,y+1\rangle\langle x,y|\otimes|11\rangle\langle 11|.
\end{aligned}
\end{equation}
 The single non-Hermitian defect  $D_2$ is defined as
\begin{equation}
D_2=\sum_{x\neq0}^{y\neq0} |x,y\rangle\langle x,y| \otimes \mathbb{I}_4 + |0,0\rangle\langle 0,0| \otimes \begin{pmatrix} e^{\gamma} & 0 &0&0\\ 0 & e^{-\gamma}&0&0\\
0&0&e^{\gamma}&0\\
0&0&0&e^{-\gamma}
\end{pmatrix}.
\end{equation}
For our 2D QW, the system size is $L_x\times L_y$, where we  set $L_x=L_y=L$.  It
follows that $U_2$ has a similar energy spectrum in Fig.~\ref{fig:4}(a).
We therefore choose a similar parameter region to reveal the scale-free localization phenomenon.

As illustrated in Fig.~\ref{fig:4}(b), the IPR of a scale-free LS scales linearly with $1/L^2$.
In Fig.~\ref{fig:4}(c), we plot the normalized spatial profiles of the eigenstates with the fifth-largest imaginary part of the eigenvalues for different system sizes at $\gamma = 3.5$. The horizontal axes are normalized by $L$. To provide a clearer observation, as shown in Fig.~\ref{fig:4}(d), we plot the spatial profiles on a logarithmic scale. Regardless of the system size, these profiles almost overlap, further demonstrating that these states are scale-free LSs.

\textit {Conclusion.}---In this Letter, we  unveil that scale-free localization can be induced  by a single non-Hermitian defect in QWs.
In fact, a single non-Hermitian defect induces effects that not only occur in various non-Hermitian systems but also interact with other non-Hermitian phenomena, as discussed in Refs.~\cite{KHLKL23,LSYL24}.
More interestingly, the emergence of scale-free localization is deeply connected to the breaking of PH-symmetry in the 1D QW.
In addition, a further increase in the non-Hermitian strength induces another phase transition, giving rise to two additional pairs of bound states.
 Furthermore, by analyzing the spectral and localization properties, we identify the critical points of the two phase transitions, which correspond to the PH-symmetry breaking transition and the bound-state emergence transition, respectively.
Moreover, we extend
our investigation to 2D QWs, where a single non-Hermitian defect also induces scale-free localization, thereby demonstrating the universality of this phenomenon.
Our results highlight the characteristics of scale-free LSs and offer a new perspective for studying non-Hermitian defect physics, thereby paving the way for future explorations in higher-dimensional QWs. 
Meanwhile, we expect that our findings will open a promising avenue for the experimental investigation of scale-free localization.

\textit {Acknowledgments.---}This work is supported by the National Key R\&D Program of China (Grant No.
	2023YFA1406701) and National Natural Science Foundation of China (Grants
	No. 92265209, 12474352, and 92476106). L. Xiao acknowledges support from Beijing National Laboratory for
	Condensed Matter Physics (No. 2024BNLCMPKF010). C. X. Guo acknowledges support from the National Natural Science Foundation of China (Grant No. 12405030) and Beijing National Laboratory for Condensed Matter Physics (No. 2025BNLCMPKF023).

\textit{Data availability.---}The data supporting the findings of this study are available from the corresponding authors upon reasonable request.

\end{document}